\documentclass[preprint,epsf,psfig,nofootinbib]{revtex4}

\usepackage{amsmath}
\usepackage{amssymb}
\usepackage{amsfonts}
\usepackage{amscd}

\usepackage[pdftex,colorlinks=true]{hyperref}
\usepackage{bm}
\usepackage{latexsym}
\usepackage[latin1]{inputenc}
\usepackage{graphicx}
\usepackage{subfigure}

\usepackage{epsfig}

\begin{document}
\title{Banishing AdS ghosts with a UV cutoff}
\author{  Tom\'as Andrade}
\affiliation{Department of Physics, UCSB, Santa Barbara, CA 93106, USA}
\author{Thomas Faulkner}
\affiliation{KITP, Santa Barbara, CA 93106, USA}
\author{Donald Marolf}
\affiliation{Department of Physics, UCSB, Santa Barbara, CA 93106, USA}
\preprint{ NSF-KITP-11-267}

\begin{abstract}
A recent attempt to make sense of scalars in AdS with ``Neumann 
boundary conditions'' outside of the usual BF-window  $-(d/2)^2 < m^2 l^2 < -(d/2)^2 + 1$ led to pathologies including (depending on the precise context) either  IR divergences  or the appearance of ghosts.  Here we argue that such ghosts may be banished by imposing a UV cutoff.  It is also possible to achieve this goal in certain UV completions.  An example is the above AdS theory with a radial cutoff supplemented by particular boundary conditions on the cutoff surface.   In this case we explicitly identify a region of parameter space for which the theory is ghost free. At low energies, this theory may be interpreted as the standard dual CFT (defined with ``Dirichlet'' boundary conditions) interacting with an extra scalar via an irrelevant interaction.   We also discuss the relationship to recent works on holographic fermi surfaces and quantum criticality.
\end{abstract}

\maketitle

\section{Introduction}

AdS/CFT relates a set of Conformal Field Theories to gravitational
theories in AdS \cite{Maldacena:1997re,Witten:1998qj,Gubser:1998bc}.
Interesting field theory dynamics follows from simple relevant deformations of these CFTs.  The inclusion of multi-trace deformations has lead to many
results \cite{Witten:2001ua,Berkooz:2002ug,Klebanov:1999tb,Gubser:2002vv,Marolf:2006nd}, and in particular to recent attempts to drive a theory across a quantum
phase transition \cite{Faulkner:2010fh,Faulkner:2010gj,Jensen:2011af,Iqbal:2011aj}. In addition, the
role of multi-trace deformations  in the holographic renormalization group has
recently been emphasized in \cite{Heemskerk:2010hk,Faulkner:2010jy} (see also \cite{Vecchi:2010dd}).  As a result, one would like to have as complete an understanding as possible of which multi-trace
deformations are allowed, and when they can lead to useful dynamics.

Linear scalars in AdS offer a good starting point for such analyses.
Within the BF window $-(d/2)^2 < m^2l^2 < -(d/2)^2 +1$ there are two possible boundary
conditions preserving conformal invariance \cite{BF82}, often called the standard and alternate quantizations
\cite{Klebanov:1999tb}. These fixed points are characterized by the existence of a
single trace operator with dimensions $d/2 + \nu$ and $d/2 - \nu$ respectively, where
$\nu^2 = m^2 l^2 + (d/2)^2$.  From the bulk perspective, it is natural to think of these as generalized Dirichlet and Neumann boundary conditions
respectively.  There are many other boundary conditions that do not preserve conformal invariance but
which correspond to multi-trace deformations of the aforementioned
choices \cite{Witten:2001ua,Berkooz:2002ug}.
For example, when it is relevant, the double trace deformation leads to an RG flow between the alternative and standard theories with the former being a UV fixed point and the later an IR fixed point.

The obstruction to playing these games for $\nu > 1$ (outside the BF window) is
that the existence of the alternative fixed point would require  an operator whose
dimension is below the unitarity bound (i.e., $d/2 - \nu < d/2 -1$). It is thus natural to suppose that only the
standard fixed point exists in this regime.  The details were studied in
\cite{Andrade:2011dg} which largely confirmed this picture, though it should be remarked that the exact issued identified at the supposed alternative fixed point was not the existence of a ghost but, instead, an IR divergence and an associated null mode. Nevertheless, ghosts do appear when this theory is deformed in various ways, including in both choosing the boundary metric to enact an IR cut-off\footnote{By taking it to be a cylinder \cite{Andrade:2011dg}, de Siter space, or anti-de Sitter space \cite{Andrade:2011nh}.} and the addition of double-trace operators.  It is thus hard to make sense of this fixed point and, indeed, at first glance it may also seem hard to make sense of double trace deformations of the standard fixed point.

This result seems at odds with recent discussions of holographic fermi surfaces
and quantum criticality \cite{Faulkner:2010tq,Faulkner:2010gj}.
The reason for concern can be abstracted as follows to the setting
of a scalar in AdS.  Suppose that we couple the CFT in standard quantization
to a propagating boundary scalar field. Consider:
\begin{equation}
\label{eq:couple}
S' = S_{CFT}^{(std)} + \frac{1}{2} \int d^dx \left(- \kappa \left(\partial \Phi\right )^2
-  \lambda \Phi^2  + \ldots \right) + S_{int} \ ,
\, \qquad S_{int} = g  \int d^dx \mathcal{\hat{O}} \Phi
\end{equation}
where $S_{CFT}$ denotes the action of the dual CFT which contains an
operator $\hat{O}$ of dimension $\Delta = d/2 + \nu$.
Note that the BF window corresponds to $0 < \nu < 1$. The free operator
dimension of $\Phi$ is $(d-2)/2$ (based on power counting using a canonical kinetic term)
from which the dimension of $g$ is $1-\nu$; thus
the interaction term is relevant for $0 < \nu < 1$. In this case we can ignore the kinetic term in the IR
and integrate out $\Phi$ (treating it as non-propagating).
This fixes $\Phi = (g/\lambda) \mathcal{\hat{O}}$ and upon substitution
in the action one finds
\begin{equation}
S' = S_{CFT}^{(std)}  + \int d^d x \frac{g^2}{2\lambda}  \mathcal{\hat{O}}^2,
\end{equation}
which is just
a double trace deformation of the CFT in standard quantization.
Furthermore, sending $\lambda  \rightarrow 0$ corresponds
to the UV fixed point, which is the alternative quantization.  That is,
for $0 < \nu < 1$ we may construct the alternate quantization from a good theory by  starting with (\ref{eq:couple}), dropping the kinetic terms (setting $\kappa = 0$)
and sending $\lambda \rightarrow 0$. The mode $\Phi$, which is
being integrated over, plays the role of an operator in $S_{CFT}^{(alt)}$
with dimension $d/2 - \nu$ (from power counting with respect to the coupling
term after setting $[g] = 0$). It also enacts the Legendre transformation which
relates the two theories \cite{Klebanov:1999tb}.

Let us attempt to continue these arguments
to $\nu>1$. It is no longer valid to integrate out $\Phi$ due to the
importance of the kinetic terms. Indeed, since $g \rightarrow 0$ in the IR
we find the low energy theory is a CFT decoupled from a free scalar field.
The regime where one might expect to obtain a good theory
is $\kappa > 0$ (so that the decoupled scalar sector is ghost free).
On the other hand, any $\lambda$ should be allowed as
$\lambda < 0$ simply induces condensation of the field $\left<\Phi\right> \neq 0$ without pathology, at least so long as appropriate higher order interaction terms (such as $\Phi^4$) are present\footnote{Note that this condensation will have a residual effect on the CFT through an irrelevant interaction.}.

Unfortunately, it turns out that ghosts appear even for $\kappa > 0$ \cite{Andrade:2011dg}.  This may be seen by examining the two-point function of $\Phi$ which, using large-$N$ factorization (see for example \cite{Faulkner:2010tq}), is given by
\begin{equation}
\label{eq:semi}
G_{\Phi}(p) = \frac{ 1}{ - \kappa p^2 - \lambda - g^2 G_\mathcal{O} (p) } .
\end{equation}
Here $G_{\mathcal{O}}$ is the two point function of $\mathcal{O}$ in the
interacting CFT.  Conformal invariance fixes  $G_\mathcal{O} = c_\nu (p^2)^ \nu$
(where $p^2  =  - \omega^2 + \vec{p}^2$) and the condition that the spectral density ${\rm Im} \ G_\mathcal{O} (\omega + i\epsilon, \vec{p})$ be positive for $\omega > 0$ further requires $c_\nu \sin(\pi \nu) > 0$.  Let us now examine $G_\Phi$ for potential ghosts. For simplicity, we restrict to the case $1 < \nu < 2$ (where $c_{\nu} < 0$), though we expect similar results for larger $\nu$.    This case was studied explicitly in \cite{Andrade:2011dg}, which showed that ghosts arise for all values of $\lambda$ and $\kappa$ (though we discuss only $\kappa >0 $ here). For $\lambda > 0$ there is always a tachyonic pole with $p_\star^2 >0$.
Expanding $G_\Phi(p \approx p_\star)$ around this pole one can show that it
has a negative residue. For $\lambda < 0$ there are now two poles, which
for large enough $\lambda$ merge and move into the complex plane.  In the real case one of these two poles is a ghost while the other is a non-ghost tachyon.  As usual, the complex case necessarily contains a ghost.

On the other hand, it is clear that no ghost is present for $g=0.$  Studying the change in the corresponding pole perturbatively in $g$ would not have indicated the presence of ghosts.  This suggests that the ghosts correspond to new poles that enter from $p=\infty$ and thus that, at least in some sense, they are a UV issue.   Indeed, since the coupling between the CFT and $\Phi$ is governed by
an irrelevant interaction we expect to run into problems at energy
scales above:
\begin{equation}
\label{eq:uvcut}
p > \Lambda_g = (g/\kappa^{1/2})^{1/(1-\nu)}.
\end{equation}
One can show that the ghost found using (\ref{eq:semi}) for $\kappa >0$ always satisfies
$|p_\star| > N_\nu \Lambda_g$ where\footnote{This explicit bound corresponds to the value of $p/\Lambda_g$ which maximizes the expression $\kappa p^2 + g^2 G_{\cal O}(p)$, associated with the case $\lambda =0$. } $N_\nu = (-\nu c_\nu)^{-\frac{1}{2(\nu-1)}}$ is a number
which depends only on $\nu$. So it is natural to expect that cutting off (or appropriately modifying) the theory
at $ p > N_{\nu} \Lambda_g$ will banish our ghosts.

The purpose of the present paper is to construct examples in which this can be demonstrated precisely.  But let us first comment on some related examples already known in the literature.
The low energy theory for the fermions analyzed in \cite{Lee:2008xf,Cubrovic:2009ye,Faulkner:2009wj} and identified in \cite{Faulkner:2010tq}  was given by an
action similar to (\ref{eq:couple}). The free fermion plays the role of our free scalar above, and the relativistic CFT above is replaced by the strongly coupled theory dual
to $AdS_2 \times R^2$ (or an interesting generalization thereof)\footnote{
The generalization of fermions to scalars in the extremal charged black hole
background was considered in \cite{Faulkner:2010gj,Iqbal:2011aj,Jensen:2011af} and a similar discussion applies. }. Although the details are different, there were again two interesting cases distinguished by conditions analogous to the cases
 $0 < \nu < 1$ and $\nu > 1$ above.
In the former case the fermion kinetic terms could be ignored and the
theory describes a non-Fermi liquid without a well defined quasiparticle.
In the later case the kinetic terms could not be ignored, resulting in
a low energy Fermionic quasiparticle different from, but similar to, a Landau Fermi Liquid.
For the case with a quasiparticle excitation it seems likely that
the low energy effective action in \cite{Faulkner:2010tq} leads to a propagator with ghost-poles.  However, the saving grace in this case is the presence of a natural UV cutoff.  Recall that
the full background considered in \cite{Lee:2008xf,Cubrovic:2009ye,Faulkner:2009wj}
was just the extremal Reissner Nordstrom charged black hole. This background can be thought of as a
domain wall solution between $AdS_4$ and
$AdS_2 \times R^2$, with the transition happening at an energy scale $\mu$ set by the chemical potential.
This $\mu$ provides an effective UV cutoff on the $AdS_2 \times R^2$ theory. The kinetic terms (analogous to $\kappa$) and $g$ were computed in \cite{Faulkner:2009wj} and one may check that they satisfy $\Lambda_g \gtrsim \mu$.
As a result, the above prediction of ghosts (based on analyzing the low energy action) is not reliable and one must instead consider the full RG flow.

In this way, the  action (\ref{eq:couple}) may generally be taken to model the IR regime of a domain wall flow between two different scale invariant fixed points.
The low energy theory then naturally comes with
a cutoff $\Lambda$; the scale where the domain wall begins to deviate from the IR fixed point.  So long as we start with a good theory in the UV,  we expect the {\it full} theory to be ghost-free.  But it is easy to engineer models in which the IR fixed point  has a field satisfying $\nu > 1$
(for the appropriately defined $\nu$) subject to an irrelevant double-trace deformation.  In this case our discussion above implies that the low-energy effective kinetic terms and the low energy
coupling will satisfy $\Lambda_g \gtrsim \Lambda$.

The problem of the existence of negative norm
states can be studied systematically on a case by case basis. Here we
take a much simpler approach and study the $AdS$ theory with
a radial cutoff. This problem is then a simple generalization of the analysis
in \cite{Andrade:2011dg} whose results will confirm the above intuition. This in turn increases one's confidence in the theories
studied in \cite{Faulkner:2010tq,Faulkner:2010gj,Iqbal:2011aj,Jensen:2011af}.

The plan of this paper is as follows:  In section \ref{ref} we introduce a simple `reference' system involving a free scalar on the Poincar\'e patch of AdS subject to a specific {\it radial} cut-off.  While this is not equivalent to a UV cut-off (since arbitrarily high momenta along the boundary are still allowed), it corresponds to a non-trivial (and non-local, see e.g. \cite{Faulkner:2010jy}) deformation of an appropriate dual CFT defined by removing the radial cut-off.  This theory is easy to study and ghost-free, but it is ill-defined at the quantum level due to an IR divergence in the two-point function (of the sort seen in \cite{Grinstein:2008qk}, \cite{Andrade:2011dg}).  Section \ref{def thy} then studies a two-parameter family of (quadratic) deformations of our reference theory.  It was shown in \cite{Andrade:2011dg} that, without the radial cut-off, these deformations remove the IR divergence but also introduce ghosts.  Nevertheless, we show that (at least in a certain regime of parameter space) the ghosts may be banished by imposing a suitably strong radial cut-off.  We close with some final discussion in section \ref{disc}, which in particular shows that the models of section \ref{def thy} suffice to give a ghost-free UV-modified version of all models studied in \cite{Andrade:2011dg} for which a certain UV coupling is positive.

\section{Reference System with Radial cutoff}
\label{ref}

As stated above, the explicit model that we will study is that of a scalar field $\phi$ on (Poincar\'e) $AdS_{d+1}$.  We impose a radial cut-off at some $r=r_0$ in coordinates associated with the metric
\begin{equation}\label{AdS}
    ds^2 = \frac{dr^2}{r^2} + \frac{1}{r^2}\eta_{ij} dx^i dx^j .
\end{equation}
In particular, we take  $r \in  (r_0, \infty)$ and note that $r = \infty$ is the Poincar\'e horizon. We focus on the mass range $1< \nu < 2$ .  For the moment, we analyze only a specific choice of boundary conditions discussed below.  While this will turn out to lead to an ill-defined quantum theory, it is easy to study and will be of use in section \ref{def thy} as a convenient reference system about which to deform.

To facilitate contact with the case $r_0 =0$ (no cut-off), we write the action in a form that parallels the $r_0 =0$ action for Neumann boundary conditions (see \cite{Andrade:2011dg}),
\begin{equation}\label{I N}
    I_{Ref} = I_0 + \int_{\partial M} \sqrt{\gamma} \left [\rho_\mu \partial^\mu \phi \phi - \frac{\Delta_-}{2}\phi^2 + \frac{1}{4(\nu-1)} \gamma^{ij} \partial_i \phi \partial_j \phi \right ] ,
\end{equation}
where $I_0 = - \frac{1}{2} \int_M \sqrt{g} [g^{\mu \nu} \partial_\mu \phi \partial_\nu \phi + m^2 \phi^2]$, $\Delta_- = (d/2 - \nu) $, $\partial M$ denotes the surface $r=r_0$ and $\rho_\mu$ is the unit normal to this surface (we denote the normal derivative by $\partial_\rho$ below). The boundary conditions must be chosen to make $I_{Ref}$ stationary. Varying (\ref{I N}) with respect to $\phi$ we obtain the boundary condition\footnote{The explicit variation is of the form $\delta I_{Ref} = \int_{\partial M} \phi \delta {\rm b.c.} $ so that the b.c. plays the role of a source in the dual theory.}
\begin{equation}\label{bc r0 pureN}
    \partial_\rho \phi = \Delta_- \phi + \frac{1}{2(\nu-1)} \Box_{\gamma} \phi \,\,\,\,\ {\rm at } \,\,\, r = r_0 .
\end{equation}
 Noting that $\partial_\rho \phi = r \partial_r \phi$, $\Box_\gamma \phi = r^2 \Box_0 \phi$ and that at small $r$ the field $\phi$ has the asymptotic expansion
\begin{equation}\label{asympt}
    \phi = r^{d/2-\nu} ( \phi^{(0)} + r^2 \phi^{(1)} + r^{2\nu} \phi^{(\nu)} + \ldots  )  \ \ \ \ {\rm with} \     \phi^{(1)} = \frac{1}{4(\nu-1)}\Box_0 \phi^{(0)} ,
\end{equation}
\noindent we can readily verify that (\ref{bc r0 pureN}) reduces to $\phi^{(\nu)} = 0$ in the limit $r_0 \rightarrow 0$. Here $\Box_0$ is the D'Alembertian associated with the flat boundary metric, i.e. $\Box_0 = \eta^{ij} \partial_i \partial_j$.

Using the prescription of \cite{Compere:2008us}, we can read off the inner product associated with the action \eqref{I N}, including necessary contributions from the boundary kinetic terms on $\partial M$. We take the bulk Klein-Gordon current associated with a pair of solutions $\phi_1,\phi_2$ to be
\begin{equation}
\label{KGbulk}
j^{bulk}_\mu =  \frac{i}{2} \phi_1^* \stackrel
{\leftrightarrow}
\partial_\mu \phi_2 ,
\end{equation}
and introduce a corresponding boundary current
\begin{equation}
\label{KGbndy}
j^{bndy}_j =  \frac{i}{2} \phi_1^* \stackrel {\leftrightarrow} \partial_j \phi_2 ,
\end{equation}
where $A \stackrel {\leftrightarrow} \partial B = A \partial B - B \partial A $ and the index $j$ ranges only over boundary directions.
The renormalized inner product is then simply
\begin{equation}
\label{ip}
(\phi_1,\phi_2) = (\phi_1,\phi_2)_{bulk} - \frac{1}{2(\nu-1)} (\phi_1,\phi_2)_{bndy} ,
\end{equation}
where $(\phi_1,\phi_2)_{bulk}, (\phi_1,\phi_2)_{bndy}$ are given by introducing some surface $\Sigma$ with boundary $\partial \Sigma$ at $r=r_0$, contracting the currents (\ref{KGbulk}), (\ref{KGbndy}) with either the normal $n^\mu $ to $\Sigma$ or the normal $n_\partial^\mu$ to $\partial \Sigma$ within the surface $r=r_0$, and integrating over $\Sigma$ or $\partial \Sigma$ using the volume measure induced by (\ref{AdS}).

\subsection{Spectrum}

In order to solve the wave equation, we shall use the mode decomposition
\begin{equation}\label{modes}
    \phi = e^{i k \cdot x} \psi_k(r) ,
\end{equation}
\noindent where $k^i = (\omega, k)$ and $\psi_k(r)$ is a radial profile that depends on the eigenvalue of $\Box_0$, which we will denote as the ``boundary mass", i.e. $m^2_{bndy} := - k_i k^i$. This eigenvalue may be used to classify the modes as  ($m_{bndy}^2 > 0$), light-like ($m_{bndy}^2 = 0$) and space-like or tachyonic ($m_{bndy}^2 < 0$). We will also consider the possibility of complex $m_{bndy}$, and refer to the associated modes as ``complex tachyons" below.

Let us first consider the time-like solutions. In this case, a general mode can be written
\begin{equation}\label{TL}
    \psi = \phi^{(\nu)} \psi_+ + \phi^{(0)} \psi_- ,
\end{equation}
\noindent where $\phi^{(0)}$ and $\phi^{(\nu)}$ are arbitrary constants and
\begin{equation}\label{D N solns}
    \psi_+ = C_{-\nu} r^{d/2} J_{\nu}(m_{bndy} r) \ \ \ \ \ \ \ \psi_- = C_{\nu} r^{d/2} J_{-\nu}(m_{bndy} r) ,
\end{equation}
\noindent with
\begin{equation}\label{Cnu}
    C_\nu = 2^{-\nu} \Gamma(1-\nu) m_{bndy}^\nu .
\end{equation}
\noindent Here $J_{\nu}(x)$ are Bessel functions of the first kind. The radial profiles \eqref{D N solns} oscillate rapidly near the Poincar\'e horizon and it can be shown both solutions are  plane-wave normalizable with respect to the inner product \eqref{ip}, see e.g. \cite{Andrade:2011dg}. Thus time-like modes form a continuum and exist for all values of $r_0$. The solution is completely specified by noting that the boundary condition (\ref{bc r0 pureN}) imposes a $r_0$-dependent relation between $\phi^{(0)}$ and $\phi^{(\nu)}$, whose explicit form will not be important for the moment. The norm of these modes follows from expression \eqref{ip} and can be computed by the methods of \cite{Andrade:2011dg}\footnote{Integrating by parts reduces the inner product to a sum of boundary terms at $r=r_0$ and $r=\infty$.  But a self-adjointness argument requires the result to be proportional to a Dirac delta-function, which can come only from the region near the horizon where the modes are plane-wave normalizeable.  It follows that only the asymptotics near $r=\infty$ are needed to compute the inner product.}. This quantity is positive definite for all $r_0$ and is given by
\begin{equation}\label{ip general bc}
    (\phi_1, \phi_2) = (2 \pi)^{d-1} \delta^{(d)}(k^i_1 - k^i_2) \ |\phi^{(0)}_{k_1} C_{\nu,k_1}  + e^{i\pi \nu} \phi^{(\nu)}_{k_1}  C_{-\nu,k_1}|^2 .
\end{equation}
\noindent As stated above, the coefficients $\phi^{(0)}$ and $\phi^{(\nu)}$ are related by the boundary conditions so that (\ref{ip general bc}) is fixed up to a normalization constant. Since for $r_0 \rightarrow 0$ we reproduce the boundary condition $\phi^{(\nu)} \rightarrow 0$, the UV behavior of (\ref{ip general bc}) is guaranteed to agree with the Neumann result of \cite{Andrade:2011dg}.

On the other hand, using the boundary condition to express $\phi^{(0)}$ in terms of $\phi^{(\nu)}$ for small $m_{bndy}$ one finds
\begin{eqnarray}\label{2pt IR D}
    \frac{(\phi,\phi)}{|\phi^{(\nu)}|^2} &\approx& \frac{4^\nu \Gamma(1+\nu)^2}{(2 \pi)^{1-d}} m_{bndy}^{-2\nu} + O(1),  \ \ \
    \label{2pt IR N}
    \frac{(\phi,\phi)}{|\phi^{(0)}|^2} \approx \frac{4^{-\nu} \Gamma(1-\nu)^2}{(2 \pi)^{1-d}} m_{bndy}^{2\nu} + O(1),
\end{eqnarray}
which coincide respectively with the Dirichlet and Neumann results for $r_0 = 0$ to leading order in $m_{bndy}$. As expected,  the leading small momentum behavior is not modified by the radial cut-off at $r_0$.  But the second expression in (\ref{2pt IR N}) means that our reference theory suffers from the same IR divergence in the bulk two-point function identified in \cite{Andrade:2011dg} for $r_0=0$ (this divergence also appeared in the pure CFT context in \cite{Grinstein:2008qk}).    Thus the theory is ill-defined at the quantum level.

Let us nevertheless complete the mode analysis for this theory.  We next consider the light-like modes, i.e. $m_{bndy} = 0$, whose general profile is
\begin{equation}\label{LL}
    \psi = A r^{d/2-\nu} + B r^{d/2+\nu} ,
\end{equation}
\noindent where $A$ and $B$ are arbitrary constants. The boundary condition (\ref{bc r0 pureN}) then implies $B = 0$. One can check that light-like modes \eqref{LL} with $B = 0$ are normalizable for $\nu > 1$ \cite{Andrade:2011dg}, and furthermore that (as one may expect from the above IR divergence) these modes are null directions of the inner product.

Finally, we discuss the tachyonic solutions characterized by $m^2_{bndy} := - p^2 < 0$. By convention, we restrict ourselves to ${\rm Re} \ p > 0$. With this choice, the normalizable solution at the Poincar\'e horizon is
\begin{equation}\label{tach}
    \psi_T = r^{d/2} K_{\nu} (p r) ,
\end{equation}
\noindent where $K_{\nu} (x)$ is the modified Bessel function of the second kind. The boundary condition (\ref{bc r0 pureN}) then yields $K_{\nu-2} (p r_0) = 0$ which, provided ${\rm Re} \ p > 0$, has no solutions anywhere in the complex plane \cite{abramowitz+stegun}. It follows that there are no tachyonic solutions.

To summarize, our reference theory is ill-defined at the quantum level due to an IR divergence in the two-point function.  This divergence is associated with the presence of null states (the light-like modes).  However, the theory has no negative-norm states.  One may therefore hope that a suitable IR modification will render the theory well-defined without introducing ghosts.  We exhibit a two-parameter family of such deformations in section \ref{def thy} below.

\section{Deformed theory}
\label{def thy}

We now deform the action \eqref{I N} by considering $I = I_{Ref} + I_{def}$ with
\begin{equation}\label{I def phi}
    I_{def} = - \nu \int_{\partial M} \sqrt{\gamma} r_0^{2 \nu} \left[ \frac{\kappa}{r_0^2} (\partial \phi)^2 + \lambda \phi^2  \right] ,
\end{equation}
\noindent where all the quantities are taken to be tensors with respect to $\gamma$. This parametrization of boundary couplings behaves smoothly in the limit $r_0 \rightarrow 0$ where it coincides with the usual notion of multitrace deformations (and in particular with the parametrization of \cite{Andrade:2011dg}).  As discussed in \cite{Andrade:2011dg}, in the absence of of a radial cut-off ($r_0=0$) such deformations {\it always} give rise to ghosts.  But below we will see that for any $\kappa > 0$ the ghosts may be banished by taking $r_0$ sufficiently large. Note that stationarity of the deformed action  requires the boundary condition
\begin{equation}\label{bcdef r0}
    \partial_\rho \phi - (\Delta_- + 2 \nu \lambda r_0^{2\nu} ) \phi - \left[\frac{1}{2(\nu-1)} - 2 \nu \kappa r_0^{2(\nu-1)} \right] \Box_{\gamma} \phi = 0  \,\,\,\,\ {\rm at } \,\,\, r = r_0.
\end{equation}

It should be noted that the deformation term \eqref{I def phi} contains a new boundary kinetic term, so that it modifies the boundary symplectic current. As a result, the total inner product reads
\begin{equation}
\label{ip2}
(\phi_1,\phi_2) = (\phi_1,\phi_2)_{bulk} - \left[ \frac{1}{2(\nu-1)} - 2 \nu r_0^{2(\nu -1)}\kappa \right] (\phi_1,\phi_2)_{bndy} .
\end{equation}
Below,  our main focus will be to find a region in the space of parameters $(\lambda, \kappa)$ that is ghost-free. To do so, we shall concentrate in the tachyonic modes, since, as shown in \cite{Andrade:2011dg}, time-like and light-like modes necessarily have non-negative norms for all $\kappa > 0$ (though the light-like modes become ghosts for $\kappa < 0$).  In particular, the light-like modes have strictly positive norms for all $\kappa > 0$ and the two-point function becomes IR finite.  Thus it remains only to analyze the possible tachyons.

\subsection{Existence of Tachyons}
\label{tachyons def thy}

We now study the existence of tachyonic solutions as we vary $r_0$ holding fixed $\lambda$ and $\kappa$. As above, we define $p^2 = - m_{bndy}^2 < 0$ and and take ${\rm Re} \ p > 0 $ by convention. We may then write the radial profile of the tachyonic solutions as
\begin{equation}\label{psi tach}
    \psi = r^{d/2} K_{\nu}(p r) .
\end{equation}
Introducing $q = p r_0$, the boundary condition (\ref{bcdef r0}) implies
\begin{equation}\label{bc def r0 SL}
    \frac{K_{\nu - 2}(q)}{K_{\nu}(q)} =  \hat{\kappa} +  \hat{\lambda}/q^2 ,
\end{equation}
\noindent where $\kappa_c = \frac{1}{4 \nu (\nu-1) r_0^{2(\nu-1)}}$,
   $\hat{\kappa} = \kappa/\kappa_c$, and $\hat{\lambda} = \lambda r_0^2/\kappa_c$.

\begin{figure}[htb]
\center
\subfigure[][]{
\label{fig1}
\includegraphics[width=0.35\linewidth]{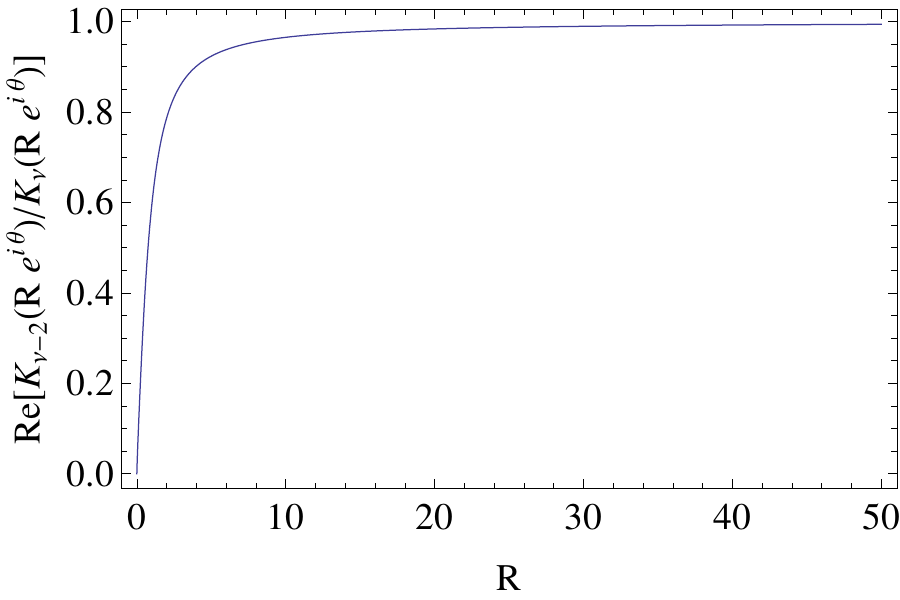}
}\qquad\qquad
\subfigure[][]{
\label{fig2}
\includegraphics[width=0.35\linewidth]{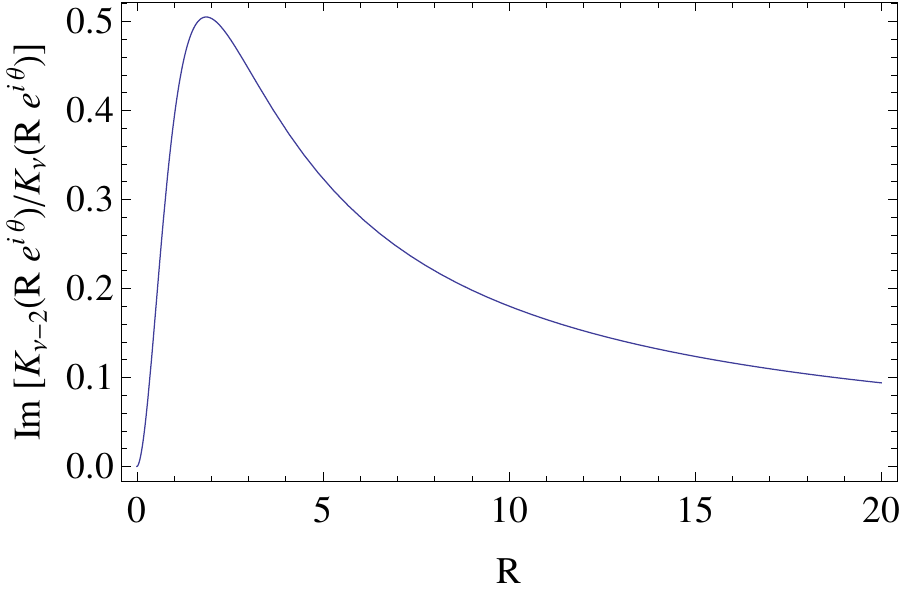}
}
\caption{On the left we plot ${\rm Re} \ \frac{K_{\nu - 2}(q)}{K_{\nu}(q)}$ vs. $R$ for $\nu = 1.4$, $\theta = 3/8 \pi$. This function is invariant under $\theta \rightarrow - \theta$. On the right we plot ${\rm Im} \ \frac{K_{\nu - 2}(q)}{K_{\nu}(q)}$ vs. $R$ for $\nu = 1.99$, $\theta = 7/16 \pi$. This function changes sign under $\theta \rightarrow - \theta$.  The peak is smaller for smaller values of $\nu$}
\end{figure}

To analyze (\ref{bc def r0 SL}), it is useful to note the following facts. First, the asymptotic form of $K_\mu(q)$ for fixed $\mu$ at large $|q|$ is
\begin{equation}\label{K large z}
    K_{\mu}(q) = \sqrt{\frac{\pi}{2q}} e^{-q} \left[ 1 + \frac{4 \mu^2 - 1}{8 q} + O(|q|^{-2}) \right ]   .
\end{equation}
Hence, letting $q = R e^{i \theta}$ we have for large $R$
\begin{equation}\label{KoverK large R}
    \frac{K_{\nu - 2}(q)}{K_{\nu}(q)} \approx 1 + \frac{2(1-\nu)}{R}( \cos \theta - i \sin \theta) + O(R^{-2}).
\end{equation}
Second, for $q \approx 0$ and ${\rm Re} \ \mu > 0$, we have $K_\mu(q) \approx \frac{1}{2} \Gamma(\mu)(\frac{1}{2} q)^{-\mu}$. In order to use this expression for $\nu -2 < 0$,  we note that $K_{-\mu}(q) = K_{\mu}(q)$. It follows that for small $R$ we can write
\begin{equation}\label{small R lhs}
    \frac{K_{\nu - 2}(z)}{K_{\nu}(z)} \approx 2^{2(1-\nu)} \frac{\Gamma(2-\nu)}{\Gamma(\nu)} R^{2(\nu-1)} \{ \cos[2(\nu-1) \theta] + i \sin[2(\nu-1) \theta]  \}.
\end{equation}
The behavior of the real and imaginary parts of the ratio of the two relevant Bessel functions is plotted in figures \ref{fig1} and \ref{fig2}. With these observations in mind, let us go back to (\ref{bc def r0 SL}).

Let us first show that there are no tachyons at complex momenta for $\lambda > 0$. To do so, we let $q = R e^{i \theta} $ with $|\theta| < \pi/2$, so (\ref{bc def r0 SL}) reads
\begin{equation}\label{Re tach eq}
    {\rm Re}
     \ \frac{K_{\nu - 2}(q)}{K_{\nu}(q)} = \hat{\kappa} + \frac{\hat{\lambda}}{R^2} \cos(2 \theta) ,
\end{equation}
\begin{equation}\label{Im tach eq}
    {\rm Im} \ \frac{K_{\nu - 2}(q)}{K_{\nu}(q)} = - \frac{\hat{\lambda}}{R^2} \sin(2 \theta) .
\end{equation}
Now, using (\ref{KoverK large R}) and the fact -- justified by numerics -- that ${\rm Im} \ \frac{K_{\nu - 2}(q)}{K_{\nu}(q)}$ has no zeroes or poles for ${\rm Re} \ q > 0$, we conclude that ${\rm Im} \ \frac{K_{\nu - 2}(q)}{K_{\nu}(q)}$ is bounded and positive definite for $0 < \theta < \pi/2$ and negative definite for $-\pi/2 < \theta < 0$. For $\hat{\lambda} > 0$ and $\theta \neq 0$, the left and right hand side of (\ref{Im tach eq}) have different signs for all $R$.  Thus there are no complex solutions.

Consider now $q \in \mathbb{R}$. It is not hard to show that\footnote{This involves using the above expansions to evaluate the LHS of (\ref{bc def r0 SL}) at large and small real $z>0$ and also showing that it is monotonic.  Monotonicity follows from positivity of the Wronskian-like quantity $W_{\nu_1,\nu_2} = z(K_{\nu_2} \partial_z K_{\nu_1} - K_{\nu_1} \partial_z K_{\nu_2} )$ for $\nu_1 < \nu_2.$  To show positivity of
$W_{\nu_1,\nu_2}$, one uses the Bessel equation to show that $W_{\nu_1,\nu_2}$ is strictly decreasing for $\nu_1 < \nu_2$ and real $z > 0$.  The argument is completed by noting that  (\ref{K large z}) implies $W_{\nu_1,\nu_2} > 0$ for large $z$.} the left hand side of (\ref{bc def r0 SL}) ranges monotonically over $(0,1)$ as $q$ varies between $(0,\infty)$.
Thus, for $\lambda > 0$, it follows that (\ref{bc def r0 SL}) has one and only one real solution if $\hat{\kappa} < 1$ (or equivalently, $\kappa < \kappa_c$) and no solutions otherwise. Recalling the definition of $\kappa_c$, we conclude that for $\lambda > 0$, $\kappa >0$ the spectrum will be tachyon free when $r_0$ is sufficiently large. Thus, at least in this regime, the resulting theories are both well-defined and ghost-free.

To make contact with the introduction note that the
the condition for a ghost free spectrum $\hat{\kappa} > 1$ can be written as:
\begin{equation}
r_0^{-1} \lesssim  \kappa^{1/2(\nu-1)} \equiv \Lambda_g
\end{equation}
where we have appropriately set $g=1$ in the expression (\ref{eq:uvcut}) for $\Lambda_g$.
So as long as the cutoff energy scale $r_0^{-1}$ is smaller than $\Lambda_g$
the theory is ghost free.
\\

\subsection{Complete Analysis}

For completeness, we now analyze the case $\lambda < 0$ and also compute the norms of the tachyons (for both signs of $\lambda$). Though our arguments above were largely analytic, we rely on simple numerics below to establish some general trends.

\begin{figure}[htb]
\center
\subfigure[][]{
\label{blobs}
\includegraphics[width=0.35\linewidth]{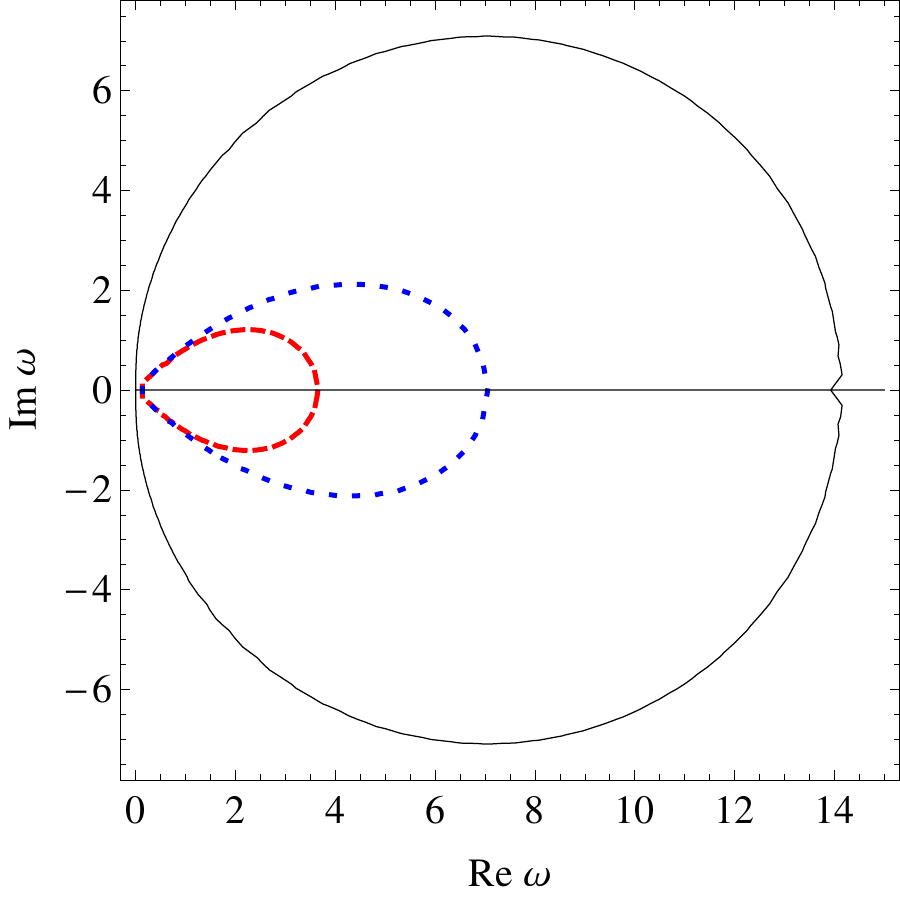}
}\qquad\qquad
\subfigure[][]{
\label{blobs2}
\includegraphics[width=0.35\linewidth]{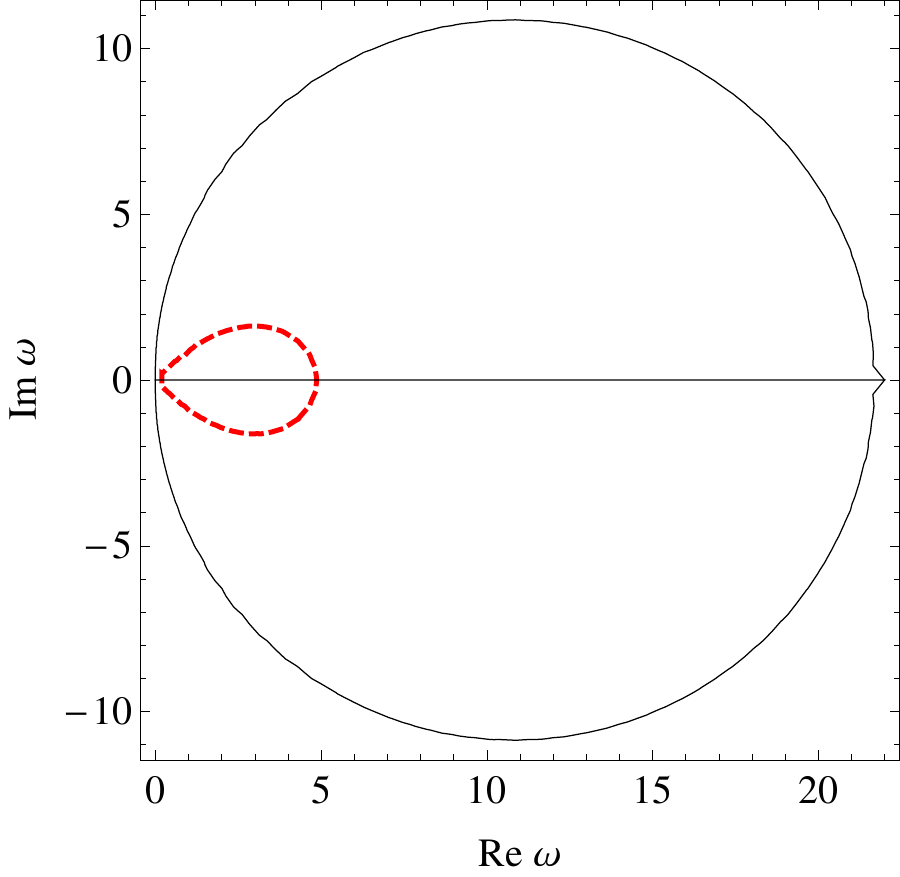}
}
\caption{The case $\hat \kappa \ge 1, \hat \lambda < 0$.  We plot
numerical solutions of (\ref{Re tach eq}) -- dashed and dotted lines -- and (\ref{Im tach eq}) -- solid lines, including both the straight lines along the real axes and the rough circles.
A simultaneous solution to both equations would requires these curves to intersect.  Since the intersection at $q=0$ corresponds to the light-like modes already studied (and is not a tachyon), there is a single real tachyon in each case shown.
Figure (a) shows results for $\hat{\lambda} = - 5$,  $\nu = 1.4$. Note that (\ref{Im tach eq}) is independent of $\hat \kappa$.  For (\ref{Im tach eq}) we show $\hat{\kappa} = 1.2$ (dashed curve) and $\hat \kappa = 1$ (dotted curve).
Figure (b) shows results for $\hat{\lambda} = - 8$, $\hat{\kappa} = 1.2$, $\nu = 1.4$. The structure is similar for all $\hat \kappa \ge 1, \hat \lambda < 0$.}
\end{figure}

\begin{figure}[htb]
\center
\subfigure[][]{
\label{cplex solns}
\includegraphics[width=0.35\linewidth]{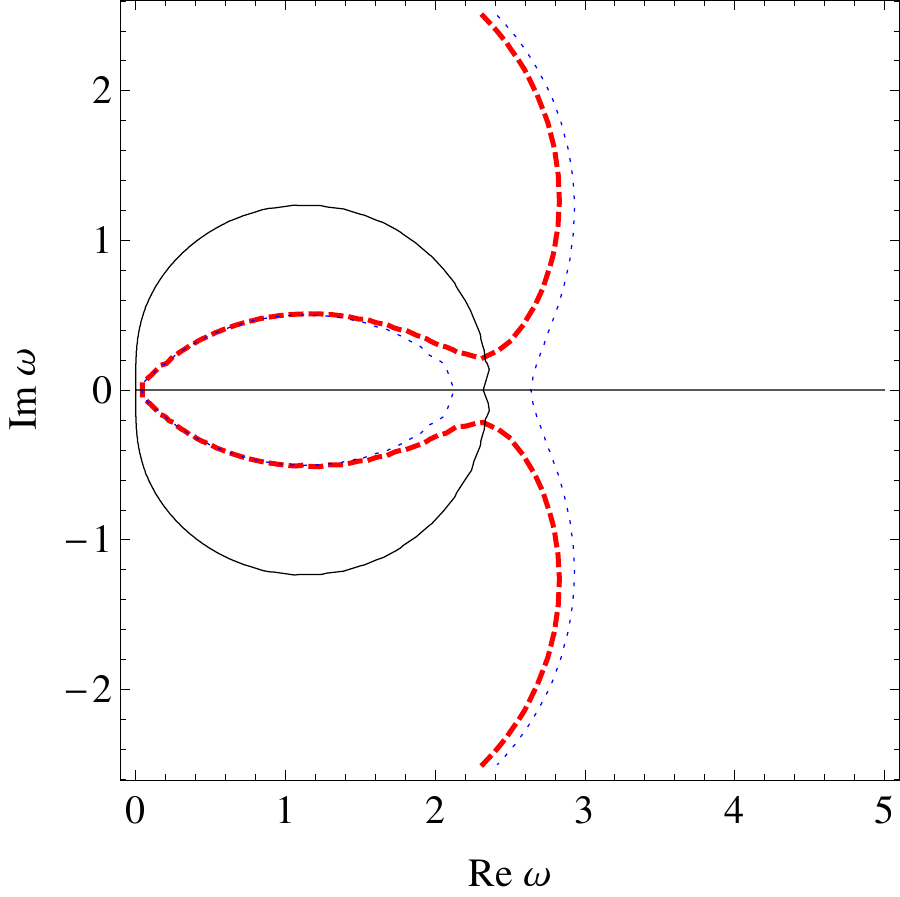}
}\qquad\qquad
\subfigure[][]{
\label{cplex solns2}
\includegraphics[width=0.35\linewidth]{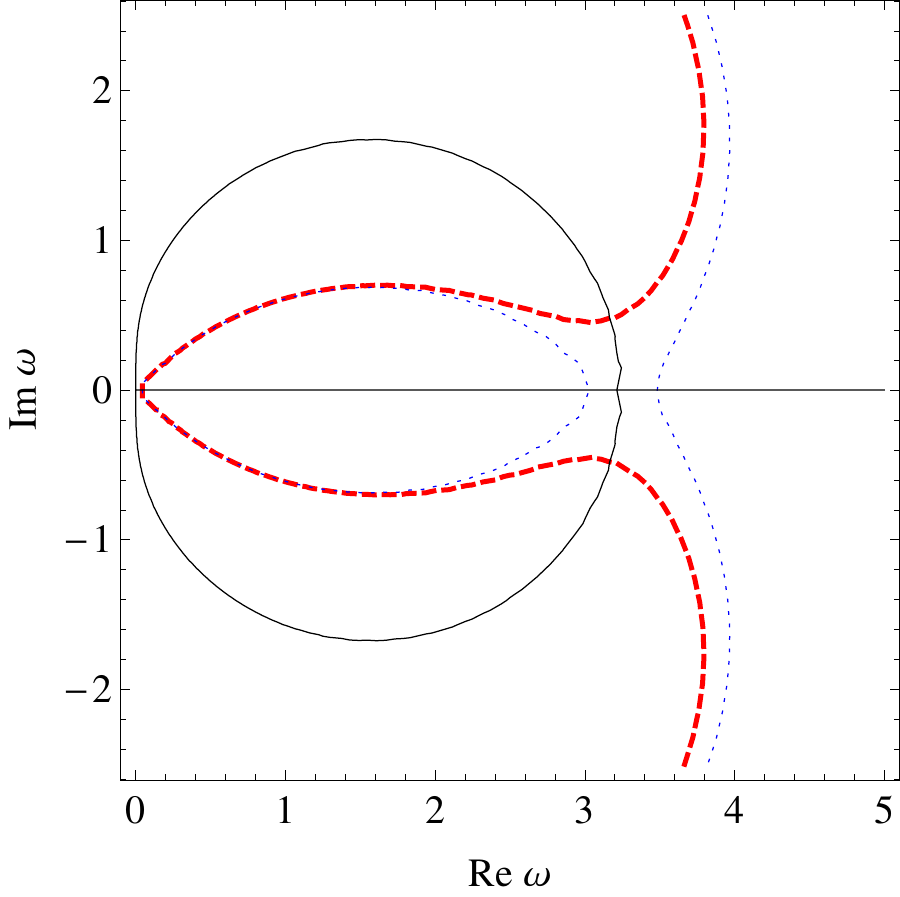}
}
\caption{
The case $\hat \kappa \le 1, \hat \lambda < 0$.  We again plot
numerical solutions of (\ref{Re tach eq}) -- dashed and dotted lines -- and (\ref{Im tach eq}) -- solid lines, including both the straight lines along the real axes and the rough circles.
Simultaneous solutions occur at the intersections. Again, $q=0$ corresponds to the light-like modes already studied (and is not a tachyon).  Figure (a) shows results for
$\hat{\lambda} = - 0.5$ and  $\nu = 1.4$.  Solutions of (\ref{Im tach eq}) are independent of $\hat \kappa$, while for (\ref{Re tach eq}) the dashed and dotted curves respectively describe $\hat{\kappa} = 0.842, 0.844$.  As suggested by the plot, increasing $\hat \kappa$ through this range causes the dashed curve to pinch off and to separate into two pieces (as shown by the dotted curves).  Further increasing $\hat{\kappa} \rightarrow 1$, the rightmost dotted line moves off to infinity and we recover figures \ref{blobs}) and \ref{blobs2}.  Changing $\hat \lambda$ appears to simply change the overall scale of the figures as indicated by figure (b) which shows  $\hat{\lambda} = - 0.8$, $\nu = 1.4$, and  $\hat{\kappa} = 0.880, 0.883$.}
\end{figure}

We begin with the case $\lambda < 0$, $\hat \kappa > 1$. For real $q$, it is easy to see that there is one real tachyon (at some positive $q$).   But numerical investigation shows that there are no complex solutions; see figure \ref{blobs}.  On the other hand, due to a new branch of solutions to (\ref{Re tach eq}) that comes in from infinty at $\hat \kappa =1$, for $\hat{\kappa} < 1$ we find either two real or two complex solutions depending on the ratio $\hat{\kappa}/\hat{\lambda}$. See figure \ref{cplex solns}.

It remains to compute the norms of the tachyonic solutions for both $\lambda > 0$ and $\lambda < 0$ (in cases where they exist). Now, the norm of a complex momentum tachyon necessarily vanishes due to symmetries.  However, since all parameters are real, the momenta of the complex tachyons come in pairs $p_\star, p_\star^*$. The inner product $(\psi(p_\star), \psi(p_\star^*))$ is non-zero, and diagonalizing the the resulting symplectic structure gives one degree of freedom with positive norm and a second with negative norm.  Thus complex tachyons are necessarily associated with ghosts and it remains only to analyze real tachyons.

Following \cite{Andrade:2011dg} we find that for tachyonic solutions of real momentum, the inner product \eqref{ip2} simplifies to
\begin{equation}\label{ip tach general}
    (\phi_1, \phi_2) = \frac{1}{2}(\omega_1 + \omega_2) (2 \pi)^{d-1} \delta^{(d-1)}(\vec{k_1}-\vec{k_2}) e^{it(\omega_1 - \omega_2)} \langle \psi_1, \psi_2 \rangle_{SL} .
\end{equation}
Here $\langle \cdot, \cdot \rangle_{SL}$ is a Sturm-Liouville-like product with an explicit boundary contribution:
\begin{equation}\label{SL norm tach}
    \langle \psi_1, \psi_2 \rangle_{SL} = \langle \psi_1, \psi_2 \rangle_{bulk} + \langle \psi_1, \psi_2 \rangle_{bndy} ,
\end{equation}
\noindent where
\begin{equation}\label{W bulk tach}
    \langle \psi_1, \psi_2 \rangle_{bulk} = - \frac{r^{1-d}_0}{p_1^2 - p_2^2} (\psi_1 \psi_2' - \psi_2 \psi'_1) \big|_{r=r_0}^{r= \infty} ,
\end{equation}
\begin{equation}\label{W bndy tach}
    \langle \psi_1, \psi_2 \rangle_{bndy} = r_0^{2-d} \left [- \frac{1}{2(\nu-1)} + 2 \nu r_0^{2(\nu-1)} \kappa \right] \psi_1 \psi_2 .
\end{equation}
Note that (\ref{W bulk tach}) is singular when evaluated in tachyonic solutions that satisfy the boundary conditions since this fixes a particular value of $p$. In order to evaluate (\ref{W bulk tach}) for a mode with momentum $p_0$ which lies in the discrete part of the spectrum, we consider two solutions with momenta $p_1$ and $p_2$ which do not satisfy the boundary conditions, take the limit $p_1, p_2 \rightarrow p$, and impose the boundary condition that sets $p = p_0$ at the end. Applying this procedure to (\ref{W bulk tach}) and taking into account the contribution (\ref{W bndy tach}) we obtain
\begin{equation}\label{ip tach general2}
    (\phi_1, \phi_2) = (2 \pi)^{d-1} \omega_1 \delta^{(d-1)}(\vec{k_1}-\vec{k_2}) \delta_{p_1, p_2} \langle \psi_1, \psi_1 \rangle_{SL},
\end{equation}
\begin{eqnarray}\label{norm tach 2}
   \langle \psi, \psi \rangle_{SL} = A \left \{  (\hat{\kappa} - 1) + (\nu - 1) \left[ \frac{K_{\nu-1}(q) K_{\nu+1}(q)}{K_{\nu}(q)^2} -1 \right] \right\},
\end{eqnarray}
\noindent where $q$ is given implicitly by (\ref{bc def r0 SL}) and $A$ is the positive quantity
\begin{equation}\label{A}
    A =  2\frac{p^{2 \nu} r_0^2 K_\nu(q)^2}{4^{\nu} (\nu-1)\Gamma(\nu)^2}.
\end{equation}
Numerical results indicate that
the second term in (\ref{norm tach 2}) (including the factor of $\nu-1$) is positive for real $q$ and decays monotonically from $1$ to $0$ as $q$ ranges over $(0, \infty)$.  The Kronecker delta in (\ref{ip tach general2}) reflects the facts that the tachyonic spectrum is discrete and that the SL product (\ref{norm tach 2}) vanishes when $p_1 \neq p_2$. Noting that (\ref{ip tach general}) also vanishes for $\vec{k}_1 \neq \vec{k}_2$, we conclude that the frequencies must also be equal in order for (\ref{norm tach 2}) to be non-zero. Thus  the time-dependent exponentials in (\ref{ip tach general}) cancel, making manifest that the inner product is conserved.  As a consistency check, we note that taking the limit $r_0 \rightarrow 0$ in (\ref{norm tach 2}) reproduces the result of \cite{Andrade:2011dg}, i.e.
\begin{equation}\label{UV limit tach norm}
    \langle \psi, \psi \rangle_{SL} = -2 \nu\left[ \kappa (\nu-1) + \frac{\lambda \nu}{p^2} \right] + O(r_0) .
\end{equation}

\begin{figure}[htbp]
\begin{center}
\includegraphics[scale=0.75]{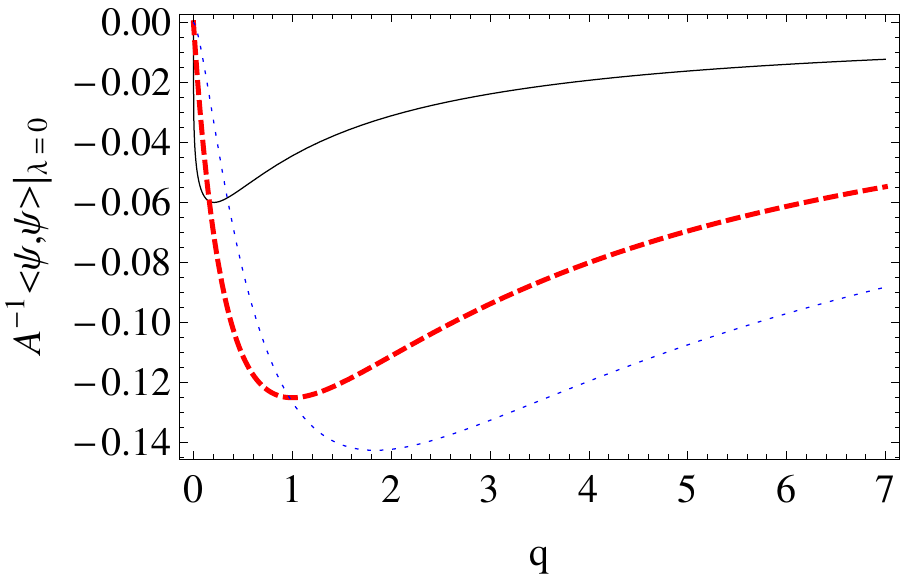}
\caption{The left hand side of \eqref{norm tach lambda 0} is plotted as a function of $q$ for $\nu = 1.1$ (solid line), $\nu = 1.5$ (dashed line), and $\nu = 1.9$ (dotted line).}
\label{norm limit}
\end{center}
\end{figure}

We now study (\ref{norm tach 2}) for the tachyons found above:

{\bf Case $\hat{\lambda} > 0$, $\hat{\kappa} < 1$}: in this region we find one real tachyon.  Since the second term in (\ref{norm tach 2}) decays monotonically, the maximum of the norm occurs when the value of $q$ that solves (\ref{bc def r0 SL}) acquires its minimum. For any fixed $\hat{\kappa} < 1$, the value of $q(\hat \lambda, \hat \kappa)$ defined by (\ref{bc def r0 SL}) decreases monotonically with $\hat{\lambda}$, arriving at the minimum when $\hat{\lambda} = 0$, see figure \ref{fig1}. Thus if the norm (\ref{norm tach 2}) is negative for $\hat{\lambda} = 0$ and all $\hat{\kappa} < 1$, it is in fact negative everywhere in the region being considered, i.e $\hat{\kappa} < 1$, $\hat{\lambda}  > 0$. To help see that this is indeed the case, we solve (\ref{bc def r0 SL}) with $\hat{\lambda} = 0$ for $\hat \kappa$ and insert the result into (\ref{norm tach 2}) to obtain:
\begin{equation}\label{norm tach lambda 0}
    \langle \psi, \psi \rangle \bigg|_{\lambda = 0} = A \left\{ K_\nu(q)^{-2} [ K_{\nu-2}(q) K_\nu(q) + (\nu-1) K_{\nu-1}(q)K_{\nu+1}(q)  ] - \nu \right\} .
\end{equation}
Plotting (\ref{norm tach lambda 0}) for $q>0$ and  $1 < \nu < 2$ shows that it is negative definite, see \ref{norm limit}.

{\bf Case $\hat{\lambda} < 0$, $\hat{\kappa} > 1$}: Here both terms in (\ref{norm tach 2}) are positive definite, so there are no ghosts.

{\bf Case $\hat{\lambda} < 0$, $\hat{\kappa} < 1$}: We found a pair of complex tachyons that can move to the real axis for certain values of $\hat{\kappa}$, $\hat{\lambda}$,  As mentioned above the complex ghosts constitute a ghost/non-ghost pair. In the region in which the tachyons are real, one may show that one (and only one) of the tachyons is a ghost by using the fact that the norm is given by the derivative of (\ref{bc def r0 SL}) up to multiplication by a positive definite function\footnote{While this may be checked explicitly using Bessel identities, it also follows from the general relation between the norm and the residues of the 2-point function.}. The norms vanish at the critical point where the tachyons leave the real axis.  At this point we expect logarithmic modes to appear with the corresponding associated ghosts.

\section{Discussion}
\label{disc}

Our main point above is that the ghosts found in \cite{Andrade:2011dg} may, at least in some cases, be banished by either imposing a suitable low UV cut-off $\Lambda_g$, or by appropriately modifying the theory on energy scales above $\Lambda_g$.  We argued that this is a general property of renormalization group flows that approach the IR fixed points of \cite{Andrade:2011dg} and which start from a well-defined UV theory, analogous to those analyzed in \cite{Lee:2008xf,Cubrovic:2009ye,Faulkner:2009wj, Faulkner:2010tq,Faulkner:2010gj,Iqbal:2011aj,Jensen:2011af}.

\begin{figure}[h]
\begin{center}
\includegraphics[scale=0.75]{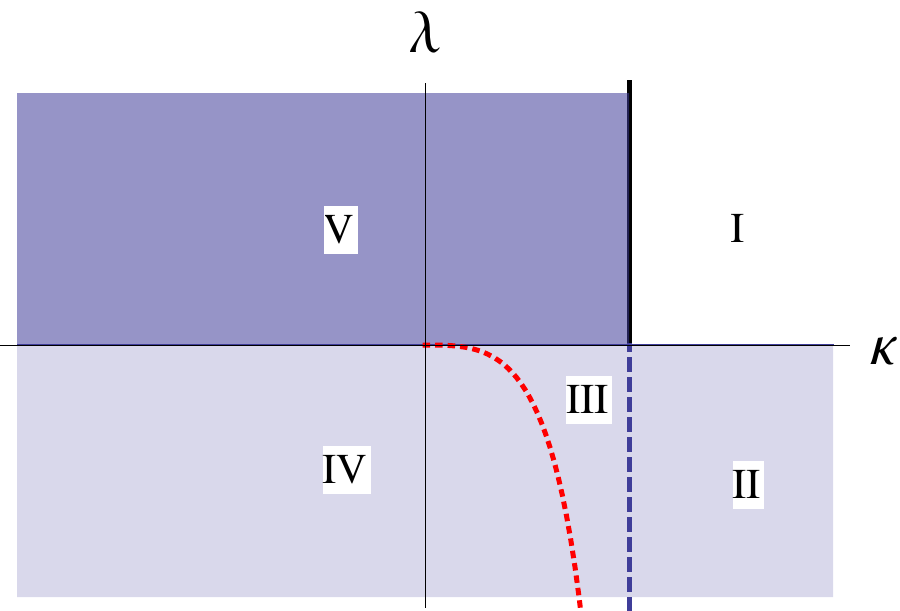}
\caption{Different regions in parameter space $(\lambda, \kappa)$}
\label{pd}
\end{center}
\end{figure}

In addition, we exhibited a simple new class of examples in which the ghosts are banished by imposing a {\it radial} cut-off on the AdS space.  As discussed in \cite{Faulkner:2010jy}, this corresponds to a non-local UV modification of the usual CFT dual to bulk AdS.  We found a two-parameter family of such theories corresponding to further quadratic deformations which are ghost-free in a certain regime of parameter space.  In particular, gathering the results found in the previous sections, leads to the phase diagram shown in figure \ref{pd}. Here, regions I and II (i.e. $\hat \kappa > 1$) constitute the ghost-free regime. More specifically, in region I there are only time-like excitations whereas in region II there is a non-ghost tachyon. On the dividing line $\lambda = 0$ a light-like mode of zero norm is present with the associated IR divergence in the 2-point function. The remaining regions contain ghosts: in region III there are two real momentum tachyons, one of which is a ghosts; in region IV there are two complex tachyons, whose presence is tied to ghosts, as explained above; finally, in region V there is one real tachyon with negative norm. Here the dotted line that marks the boundary between the regions with two real (III) and two complex tachyons (IV) is to be considered very approximate.  We have not investigated this boundary in detail, though the fact that $K_{\nu-2}(q)/K_\nu(q)$ is positive for $q > 0$ and vanishes for $q=0$ shows that it lies to the right of the $\lambda$-axis and terminates at the origin.
For small $q$ we can send the cutoff $r_0$ to zero and the boundary between
region III and IV satisfies
$\lambda \sim - \kappa^{\nu/(\nu-1)}$.

It is natural to ask whether a similar simple radial cut-off can banish more general ghosts.  Consider for example the addition of a new term to $I_{def}$ involving $\eta (\Box_\gamma \phi)^2$ (a $p^4$ term with coefficient $\eta$).  The higher order boundary condition will then give rise to additional ghosts.  Our preliminary numerical investigations indicate that for $\eta \neq 0$ there are no values of $\kappa, \lambda, \eta, r_0$ for which the theory is ghost-free, so that the ability to banish ghosts by using a simple radial cut-off is not generic.  However, it is again likely that for at least some values of the parameters that a more complicated UV modification of the IR fixed point (such as that associated with RG flow from a good UV theory) that renders the theory ghost-free.

We conclude by making explicit the sense in which the radial cut-off theories of section \ref{def thy} are UV modifications of a theory with no cut-off.  This may be done by comparing the two point functions of the theories with finite and vanishing $r_0$ in the deep IR, which we take to mean $m_{bndy}=0$.  This is in turn equivalent to studying expression (\ref{ip general bc}) for the norms at small $m_{bndy}$.  We take the cut-off free theory to be given by the same action $I = I_{Ref} + I_{def}$ with couplings $\tilde \kappa, \tilde \lambda$ and $r_0=0$.  As noted in section \ref{def thy}, our parametrization was chosen to behave smoothly as $r_0 \rightarrow 0$.

The first two leading order terms in these two-point functions agree if we identify $\hat \lambda = r_0^2 \lambda/\kappa_c$ and $\hat \kappa = \kappa/\kappa_c$ as $r_0$-dependent functions of $\tilde \lambda, \tilde \kappa$  through

\begin{equation}\label{lambda tilde}
    \hat\lambda(\tilde{\lambda}, \tilde{\kappa}) = \frac{r_0^2}{\kappa_c} \frac{\tilde{\lambda}}{1+ r_0^{2\nu} \tilde{\lambda}} \approx 4 \nu (\nu-1) ,
\end{equation}
\begin{equation}\label{kappa tilde}
    \hat\kappa(\tilde{\lambda}, \tilde{\kappa}) = \frac{[ r_0^2 \tilde{\lambda} (1+ \nu + r_0^{2 \nu} \tilde{\lambda} )  + 2 \tilde{\kappa}(\nu^2 - 1)]}{2 \kappa_c (\nu^2 - 1)(1+ r_0^{2\nu} \tilde{\lambda})^2} \approx \frac{2 \nu}{\nu + 1} ,
\end{equation}
where we have displayed the behavior for large $r_0$.
Thus we see that given any $\tilde \kappa$ and any positive\footnote{For $\tilde{\lambda} < 0$ the couplings diverge at $r_0^{-2 \nu} = -\tilde{\lambda} $, and the theories at each side of the pole are not smoothly connected as we vary $r_0$.} $\tilde \lambda $ in the $r_0=0$ theory,
for large $r_0$ the IR behavior is described by the universal values $\hat \kappa_{univ} = \frac{2 \nu}{\nu + 1}$ and $\hat \lambda_{univ} = 4 \nu (\nu-1).$  Since our analysis holds for $2 > \nu > 1$ we have
$\hat \kappa_{univ} > 1$ and $\hat \lambda_{univ} > 0$.  In this sense, subjecting such $r_0=0$ theories to a radial cut-off at large $r_0$ renders them both ghost- and tachyon-free.

The expressions (\ref{lambda tilde}) and (\ref{kappa tilde}) can be interpreted
as RG flows for the couplings $\hat{\lambda}$ and $\hat{\kappa}$
as a function of the cutoff $r_0$. Indeed they
are solutions to the RG equations of \cite{Heemskerk:2010hk,Faulkner:2010jy} where the
multi-trace couplings (or in the language of \cite{Faulkner:2010jy} the boundary action $S_B$) are truncated to second order in boundary derivatives.  The constant $\tilde{\kappa}$ and $\tilde{\lambda}$
are integration constants. Since the Wilsonian RG equations of \cite{Heemskerk:2010hk,Faulkner:2010jy} are exact, and since the spectrum does not change under exact RG, the full solutions that include
all higher derivative couplings  (but which continue to fix all other couplings to zero in the $r=0$ theory) would necessarily describe radial cut-off theories with ghosts.  In this case it is the truncation that leads
to a well-defined ghost-free theory.

\section*{Acknowledgements}
T.A. and D.M. were supported in part by the National Science Foundation under Grant No PHY08-55415,
and by funds from the University of California. This research was supported in part by the National Science Foundation under Grant No. PHY05-51164. T.A. is also pleased to thank the Department of Mathematics of the University of California, Davis, for their hospitality during the completion of this work.

\end{document}